\documentclass[%
 reprint,
showpacs,
 amsmath,amssymb,
 aps,
prl,
]{revtex4-1}

\usepackage[pdftex]{graphicx}
\usepackage{dcolumn}
\usepackage{bm}

\def\be{\begin{equation}}
\def\ee{\end{equation}}
\def\ba{\begin{eqnarray}}
\def\ea{\end{eqnarray}}

\def\rd{\mathrm{d}}
\def\rD{{\rm D}}
\def\CL{{\cal L}}
\def\g{\mathfrak{g}}
\def\Lieg{{\mathrm{Lie}(G)}}

\newcommand{\md}{\mathrm{d}}

\let\r=\rho

\newcommand*{\R}{{\mathbb R}}

\begin{document}

\title{Gauging without Initial Symmetry}

\author{Alexei Kotov}\email{oleksii.kotov@uit.no}
\affiliation{Department of Mathematics and Statistics, Faculty of Science and Technology \\ University of Troms{\o}, N-9037 Troms{\o}, Norway}

\author{Thomas Strobl}\email{strobl@math.univ-lyon1.fr}
\affiliation{Institut Camille Jordan,
Universit\'e Claude Bernard Lyon 1 \\
43 boulevard du 11 novembre 1918, 69622 Villeurbanne cedex,
France
}%
\date{April 3, 2014}
\begin{abstract}
The gauge principle is at the heart of a good part of fundamental physics: Starting with a group $G$ of so-called rigid symmetries of a functional defined over space-time $\Sigma$, the original functional is extended appropriately by additional $\mathrm{Lie}(G)$-valued 1-form gauge fields so as to lift the symmetry to $\mathrm{Maps}(\Sigma,G)$. Physically relevant quantities are then to be obtained as the quotient of the solutions to the Euler-Lagrange equations by these gauge symmetries.

 In this article we show that one can construct a gauge theory for a standard sigma model in arbitrary space-time dimensions where the target metric is not invariant with respect to \emph{any} rigid symmetry group, but satisfies a much weaker condition: It is sufficient to find a collection of vector fields $v_a$ on the target $M$
 satisfying the \emph{extended Killing equation} $v_{a(i;j)}=0$ for \emph{some} connection acting on the index $a$.  For regular foliations this is equivalent to requiring the conormal bundle to the leaves with its induced metric to be invariant under leaf-preserving diffeomorphisms of $M$, which in turn generalizes Riemannian submersions to which the notion reduces for smooth leaf spaces $M/\sim$.

The resulting gauge theory has the usual quotient effect with respect to the original ungauged theory: in this way, much more general orbits can be factored out than usually considered. In some cases these are orbits that do not correspond to an initial symmetry, but still can be  generated by a finite-dimensional Lie group $G$. Then the presented gauging procedure leads to an ordinary gauge theory with Lie algebra valued 1-form gauge fields, but showing an unconventional transformation law. In general, however, one finds that the notion of an ordinary structural Lie group is too restrictive and should be replaced by the much more general notion of a structural Lie groupoid.

\end{abstract}

\maketitle

\section{Introduction}
The Standard Model of elementary particle physics, but also General Relativity and String Theory, are gauge theories. In the former case, for example, gauging of an SU(3) rigid symmetry rotation between the three quarks leads to the introduction of the eight gluons that mediate the interaction between those elementary particles. Mathematically the resulting theory is described by connections in a principle bundle (in the above example with the structure group SU(3), the connection 1-forms representing the $\dim$ SU(3) $=8$ gluons) with the matter fields being sections in appropriate associated vector bundles.

The procedure can be generalized to matter fields being sections in
arbitrary fiber bundles, the fibers being equipped with geometric structures invariant w.r.t.~some group G, the structural or ``rigid'' symmetry group. In the case of a trivial bundle, sections correspond to maps from the base manifold $\Sigma$ to the fiber $M$ and one obtains a sigma model, such as e.g.~the ``standard'' one:
 \be \label{eq:sigma}
S_0[X] = \int\limits_\Sigma  \frac{1}{2} g_{ij}(X) \, \md X^i \wedge * \md X^j  \, .
\ee
This is a functional on smooth maps $X\colon\Sigma\to M$. The
$d$-dimensional spacetime $\Sigma$ and the  $n$-dimensional target manifold $M$ carry a (possibly Lorentzian signature) metric $h$ and $g$, respectively, $h$ entering by means of $*$.

Symmetries of the geometrical data on the source manifold $\Sigma$ or on the target manifold $M$ lift to symmetries of functionals using only such data. So, in the case of \eqref{eq:sigma} an invariance of $h$ and $g$ leads to an invariance of $S_0$. By the Noether procedure this gives rise to conserved quantities. For example, if $h$ is a flat, these conserved quantities yield the energy momentum tensor $T^{\mu\nu}$.

Let us now suppose that the metric $g$ has a nontrivial isometry group $G$, which infinitesimally implies $\CL_vg=0$,
valid for the vector fields $v=\r(\xi)$ on $M$ corresponding to arbitrary elements  $\xi\in\g=\Lieg$. In this case, there is a canonical procedure to lift the induced $G$-symmetry of $S_0$ to a gauge symmetry on an extended functional $S_1$ called \emph{minimal coupling}: After introducing $\g$-valued 1-forms $A=A^a\xi_a\in\Omega^a(M,\g)$, $\xi_a$ denoting any basis of $\g$, ordinary derivatives $\rd X^i$ on the scalar fields $X^i$ are replaced by covariant ones,
\begin{equation}\label{DX}
\rD X^i := \rd X^i - \rho_a^i(X) A^a \, ,
\end{equation}
where $\rho_a^i(x)\partial_i\equiv\rho(\xi_a)$. The new functional
\be \label{eq:sigmagauged}
S[X,A] = \int\limits_\Sigma  \frac{1}{2} g_{ij}(X) \, \rD X^i \wedge * \rD X^j  \,
\ee
is now invariant with respect to the combined infinitesimal gauge symmetries
 generated by
 \begin{eqnarray}
\delta X^i & =& \rho_a^i(X) \varepsilon^a \, , \label{deltaX} \\
\delta A^a &=& \rd \varepsilon^a + C^a_{bc} A^b \varepsilon^c \, , \label{deltaA}
\end{eqnarray}
for arbitrary $\varepsilon^a\in{C}^\infty(\Sigma)$. Here $C^a_{bc}$ are the structure constants of the Lie algebra $\g$ in the chosen basis.

In the space of (pseudo) Riemannian metrics, those permitting a non-trivial invariance or isometry group $G$ are the big exception. A generic metric $g$ does not permit \emph{any} non-vanishing vector field $v$ satisfying
${\cal{L}}_vg=0$.

It is conventional belief that an isometry is necessary to gauge the functional \eqref{eq:sigma}. It is our intention to show that this is far from true: First, there may be group actions on the target $M$ that are \emph{not isometries} but still can be gauged using Lie algebra valued 1-forms. Second, and maybe more important, one does not need to restrict to the action of finite-dimensional Lie groups. It is sufficient to have a \emph{Lie groupoid} ${\cal G}$ over $M$. In fact, the use of Lie groupoids (and their associated Lie algebroids) in the context of gauge theories is even \emph{suggested} by the present analysis as the much more generic one.

\section{The case of 1-dimensional leaves}
For a conceptual orientation, we first consider the highly simplified situation of a (regular) foliation of $M$ into \emph{one-dimensional}, hyper-surface-orthogonal leaves for a positive-definite metric $g$. In this case we can choose an adapted local coordinate system  such that $\partial_1$ generates these leaves and $x^1=const$ yields orthogonal hyper-planes. This implies that $g_{1i}=0$ for all $i\neq1$ or, if we denote those indices by Greek letters from the beginning of the alphabet, that
$g_{1\alpha}=0$ (while certainly $g_{11}>0$). $\partial_1$  \emph{not} generating an isometry is tantamount to  $g_{ij,1}\neq 0$, for at least some components.

According to standard folklore, it should not be possible to
extend $S_0$ by gauge fields such that $\partial_1$ becomes a direction of gauge symmetries, i.e.~such that $\delta{X}^1=\varepsilon$
will leave the extended action invariant for an arbitrary choice of the parameter function $\varepsilon\in{C^\infty(\Sigma)}$  (together with an appropriate transformation of the gauge field).

Since the leaves are 1-dimensional, we will introduce also only one gauge field $A\in\Omega^1(\Sigma)$ and consider the action functional
\begin{eqnarray}
 \label{eq:sigmagaugedbaby}
S[X,A] & =& \int\limits_\Sigma \frac{1}{2} g_{11}(X) \,(\rd X^1 - A) \wedge * (\rd X^1 - A) \nonumber \\ && +\frac{1}{2}g_{\alpha \beta}(X) \, \rd X^\alpha \wedge * \rd X^\beta  \, .
\end{eqnarray}
If we postulate the conventional $\delta{A}=\rd\varepsilon$, we achieve that $\rd{X^1}-A$ is strictly gauge invariant. It is then easy to see that with this transformation of the gauge field, we necessarily need $g_{ij,1}=0$, which would imply that $\partial_1$ generates isometries of $g$. However, one notices that changes of $g_{11}$ under the flow of $\partial_1$ can be compensated by means of a modified transformation of the gauge field $A$. It is sufficient to require merely $g_{\alpha \beta,1}=0$
for gauge invariance of \eqref{eq:sigmagaugedbaby} if  $\delta{X}^1=\varepsilon$ is amended by
\begin{equation}\label{eq:deltaAbaby}
\delta A = \rd \varepsilon +  \frac{\varepsilon}{2}\left(\ln(g_{11})\right)_{,1} \, (\rd X^1 - A) \, .
\end{equation}
This makes sense also geometrically: To factor out the one-dimensional leaves equipped with the metric $g_{11}$, it is not necessary that this metric, disappearing in the quotient, is invariant along the leaves, but only  that the transversal metric is:  $g_{\alpha \beta,1}=0$.

A conceptual understanding and straightforward generalization of this idea is obvious: Consider the Cartesian product $M=M_1\times{M_2}$ of two Riemannian manifolds $(M_1,g_{(1)})$ and $(M_2,g_{(2)})$ equipped with $g=g_{(1)}+g_{(2)}$, the sum of the pullback of the 2-tensors on each of the two factors by the respective projection map.
To construct a sigma model with target $(M_2,g_{(2)})$ in terms of a ``quotient construction'' for a gauge theory with target $M$,  it should not play any role that the total metric $g$ is not invariant along the leaves $M_1$. Decisive is that $g$ has this ``transversal-to-$M_1$ part'' (the pullback of $g_{(2)}$) invariant under diffeomorphisms  along the leaves $M_1$ (the invariance following precisely from the fact that it is a pullback). In fact, in \eqref{eq:sigmagaugedbaby} $g_{11}$ is not a function of only $x^1$, but it can depend on all the coordinates of $M$: Correspondingly, the  fiber-metric on $M \to M_{(2)}$ can variy along the fibers. In general we will not have to even require a fibration, but permit singular foliations.

Let us confirm our expectation that in the case of $M=\R\times{M_{(2)}}$ equipped with the adapted coordinates $(x^1,x^\alpha)$ such that the metric tensor on $M$ satisfies  $g_{\alpha \beta,1}=0$, the gauge invariant content of \eqref{eq:sigmagaugedbaby} is indeed described by a sigma model of the type \eqref{eq:sigma} with target $(M_{(2)},g_{\alpha\beta})$: Variation of  \eqref{eq:sigmagaugedbaby} w.r.t.~the gauge field $A$ leads to $g_{11}*(\rd{X^1}-A)=0$, i.e.~to $A=\rd{X^1}$, determining $A$ completely in terms of $X^1$. Since moreover $X^1$ is purely gauge by construction, neither $A$ nor $X^1$ contain any physical information. Varying the action \eqref{eq:sigmagaugedbaby} w.r.t.~the remaining fields $X^\alpha$, terms from the first line evidently vanish on-shell, while the second line gives the Euler Lagrange equations of the  expected ``reduced functional''.

In general the physical degrees of freedom cannot be separated  that easily from the unphysical ones. This is the main point of the use of gauge theories. In some sense they provide a smooth definition of an otherwise very singular quotient space.  So while  the theory has to be constructed so as to give the expected results in the simplest situations of clearly separable physical and gauge degrees of freedom, the real interest lies in those situations where this separation is either hidden or not even possible (on a global level and in a smooth manner).

\section{Arbitrary foliations}
Consider the neighborhood of a point in $M$ in which the leaves of the foliation are generated (over smooth functions) by a set of vector fields $\rho_a$, $a=1,\ldots,r$.
Clearly they must be involutive, i.e.~there will exist \emph{functions} $C^c_{ab}$ such that
\begin{equation}\label{eq:involutive}
[ \rho_a, \rho_b] = C^c_{ab} \rho_c \, .
\end{equation}
We want to promote \emph{arbitrary} deformations along the leaves
to a gauge symmetry, Equation \eqref{deltaX} with $\varepsilon^a$ arbitrary functions on $\Sigma$, at least locally. The functional \eqref{eq:sigmagaugedbaby} corresponds to a special case of \eqref{eq:sigmagauged}, only \eqref{eq:deltaAbaby} deviates from the conventional transformation behavior. It is easy to verify that, as a consequence of involutivity, also in the more general situation where  $C^c_{ab}$ are structure functions over $M$, the expressions \eqref{DX} transform covariantly, $\delta{(DX^i)}=\varepsilon^a(\rho_a^i)_{,j}\rD{X^j}$, if \eqref{deltaA} holds true.\footnote{We remark that $C^a_{bc}$ is not yet uniquely defined by means of \eqref{eq:involutive} since the set of  $\rho_a$s does not form a basis in general. At the moment we only assume a smooth, consistent choice being made.} Thus, in generalization of \eqref{eq:deltaAbaby}, we make the ansatz
\begin{equation}
\delta A^a = \rd \varepsilon^a + C^a_{bc} A^b \varepsilon^c + \Delta A^a \, . \label{deltaAneu}
\end{equation}
Here we wrote simply $(\rho_a^i)_{,j}$ and $C^a_{bc}$ for
$(\rho_a^i)_{,j}(X)\equiv{X}^*\left(\partial_i(\rho_a^i)\right)$ and $C^a_{bc}(X)\equiv{X}^*(C^a_{bc})$, respectively, and we will do likewise below.
Variation of \eqref{eq:sigmagauged} with respect to \eqref{deltaX} and \eqref{deltaAneu} yields
\begin{equation}
\delta S = \int_\Sigma \frac{1}{2} \varepsilon^a \left({\cal{L}}_{\rho_a} g\right)_{ij} \, \rD X^i \wedge * \rD X^j - g_{ij} \rho_a^i \Delta A^a \wedge * \rD X^i \, .
\end{equation}
This vanishes for all $\varepsilon^a(\sigma)$ if and only if there exist some coefficients $\omega^a_{bi}$, corresponding to an
$r\times{r}$-matrix the coefficients of which $\omega^a_{b}\equiv\omega^a_{bi}\rd{x^i}$ are locally 1-forms on $M$, such that the following holds true:
\begin{eqnarray}
\Delta A^a &:=& \omega^a_{bi} \varepsilon^b \rD X^i \label{DeltaA}\, , \\ \label{omega_conditions}
{\cal{L}}_{\rho_a} g &=& \omega^b_{a}\, \vee \, \iota_{\rho_b} g \, ,
\label{gDelta}\end{eqnarray}
where $\vee$ denotes the symmetric tensor product (for 1-forms $\alpha\vee\beta=\alpha\otimes\beta+\beta\otimes\alpha)$. It is comforting to verify that this condition is independent of the chosen generators along the leaves. For example, using a change of generating vector fields,
\begin{equation} \label{hatrho}
\widehat{\rho_a} := L^b_a \, \rho_b\,,
\end{equation}
 where $L^b_a$ are the components of an $r\times{r}$ matrix that are locally functions on $M$, this only changes the matrix $\omega$: Indeed,  \eqref{gDelta} yields
\begin{equation} \label{hatomega}
\widehat{\omega^c_a} L_c^b= L_a^c \omega_c^b + \rd L_a^b    \, .
\end{equation}

In fact, in the case of a regular foliation ${\cal{F}}$ on a Riemannian manifold $(M,g)$ there is a more geometrical formulation of the condition  \eqref{gDelta}, which is presented in the first part of the following Theorem, the second part summarizing the main findings of this letter up to this point: \vspace{-.5mm} 
 \\{\bf Theorem:} \vspace{1mm}\\
$\bullet$ Given a (regular) foliation ${\cal{F}}$ of a
Riemannian manifold $(M,g)$, then its conormal bundle $N^*\cal{F}$ 
 with its canonical metric  is invariant with respect to leaf-preserving diffeomorphisms on $M$, iff for any locally defined generating set of vector fields $(\rho_a)_{a=1}^r$ of $T{\cal{F}}$ there exists a local $r\times{r}$ matrix $\omega$ such that \eqref{gDelta} holds true or, equivalently, iff
\begin{equation} \label{decomp}
{\cal{}L}_vg\in
\Gamma({\cal{D}}^\circ\vee{}T^*M).
\end{equation}
Here ${\cal{D}}\equiv\left(T{\cal{F}}\right)^\perp\subset{}TM$ is the $g$-orthogonal complement to the tangent distribution $T\cal{F}$, $TM=T\cal{F}\oplus{}{\cal{D}}$, and  ${\cal{D}}^\circ\subset{}T^*M$ the annihilator subbundle of ${\cal{D}}$ inside $T^*M$.
\vspace{-1mm}\\$\bullet$ Let $(M,g)$ be a pseudo-Riemannian manifold and ${\cal{F}}$ a possibly singular foliation of $M$.  Suppose that for any local choice of  $(\rho_a)_{a=1}^r$ and $C_{ab}^c$ satisfying Eq.~\eqref{eq:involutive} there exists an $\omega^a_b$ such that \eqref{gDelta} holds true. Then the action functional
\eqref{eq:sigmagauged} is gauge invariant with respect to the infinitesimal gauge transformations generated by $\delta{X}^i=\rho_a^i\varepsilon^a$ and $\delta{A}^a=\rd\varepsilon^a+C^a_{bc}A^b\varepsilon^c+\omega^a_{bi}\varepsilon^b(\rd{X}^i-\rho^i_c A^c)$.
 \vspace{3mm} \\
The conormal bundle  $N^*\cal{F}$ is the annihilator-subbundle $(T{\cal{F}})^\circ\subset{}T^*M$ inheriting a fiber-metric as a subspace of  $T^*M\stackrel{g}{\approx}TM$. We did not restrict to merely 
foliation-preserving diffeomorphisms, but those which preserve \emph{each leaf} of the foliation separately. 
\vspace{4mm}\\
{\bf Proof:} Since $T^*M={\cal{D}}^\circ\oplus{}N^*{\cal{F}}$, any symmetric $2-$tensor field can be uniquely decomposed into three parts, belonging to the space of sections of $S^2{\cal{D}}^\circ\equiv{}{\cal{D}}^\circ\vee{}{\cal{D}}^\circ$, $S^2N^*{\cal{F}}$, and
${\cal{D}}^\circ\vee{}N^*{\cal{F}}$. Let $g_\parallel$ and $g_\perp$ be the first and the second components of $g$ in this decomposition, respectively; thus $g=g_\parallel+g_\perp$. In general, ${\cal{}L}_vg$ has all three components. \eqref{decomp} expresses that its second component vanishes, $({\cal{L}}_vg)_\perp=0$, which clearly follows from \eqref{gDelta}; but also vice versa:  $\cal{F}$  being regular, one can always choose a smooth collection of such $1$-forms $\omega^a_b$ under the condition \eqref{decomp}.

On the other hand, $g_\perp$ equips ${\cal{D}}$ with a metric, as well as its dual ${\cal{D}}^*\cong{}N^*{\cal{F}}$. Thus, what remains to be shown is that 
for any (local) $\cal{F}$-tangent vector field $v$, $({\cal{L}}_vg)_\perp=0$ iff 
\begin{equation} \label{gperp}
{\cal L}_v g_\perp=0 \, . 
\end{equation}
But  $({\cal{L}}_vg)_\perp=\left({\cal L}_v g_\perp\right)_\perp$ since for any $v\in\Gamma(T\cal{F})$ and
 $v_1,v_2\in\Gamma({\cal D})$: 
$\left({\cal L}_vg_\parallel\right)(v_1,v_2)=v\left(g_\parallel(v_1,v_2)\right)-g_\parallel([v,v_1],v_2)-g_\parallel(v_1,[v, v_2])=0$. Using involutivity of $T{\cal{F}}$, one shows similarly ${\cal L}_v g_\perp=\left({\cal L}_v g_\perp\right)_\perp$.
$\hfill\square$\vspace{2mm}\\

Equation \eqref{gperp} shows that if the leaf space $N:=M/\sim$ is a smooth manifold, it can be endowed with a unique Riemannian metric $h$ such that the map from $(M,g)$ to $(N,h)$ is a so-called \emph{Riemannian submersion} \cite{Neill}. 

\newpage
\section{Lie groupoids versus Lie groups}
Most of the investigations of Sophus Lie were concerned with transformations  satisfying infinitesimally conditions of the form \eqref{eq:involutive}. It was a highly non-trivial step to arrive from this to the abstract notion of a group and a Lie algebra. In the latter case it amounts to first restricting to cases where the structure functions $C^c_{ab}$ in \eqref{eq:involutive} can be chosen as constant, then postulating elements $\xi_a$ having their proper own life: they are to generate an algebra structure on the vector space $\g$ spanned by the $\xi_a$s via the product relation $\xi_a\bullet\xi_b=C^c_{ab}\xi_c$.

Let us go back to the original setting of a (possibly singular) foliation with a local description such as in \eqref{eq:involutive} and let us drop the somewhat unnatural condition, which underlie implicitly the introduction of Lie groups and algebras, that there should be a \emph{global} choice of $\rho_a$s and $C^a_{bc}$ such that the latter functions are \emph{constants} over all of $M$. First, we do not require that the vector fields $(\rho_a)_{a=1}^r$ need to be defined everywhere; we content ourselves with the fact that any neighborhood of a point permits a set of such vector fields generating the given (singular) foliation, keeping, however, the number $r$ fixed. On overlaps $U_{\alpha\beta}=U_\alpha\cap{U}_\beta$ of local charts then certainly there will be $r\times{r}$ matrices $L$ such that an equation of the type \eqref{hatrho} holds true. If the $\rho_a$s form a basis at each point of a given neighborhood, then the matrices $L$ are even unique
  (and satisfy automatically a cocycle condition). In general, however, the generating vector fields may be linearly dependent, and thus the transition matrices $L$ not uniquely determined. Let us assume that also in this case there is a consistent choice of these matrices such that they are invertible on each overlap and that on triple overlaps they fit together consistently.

Such a consistent choice of matrices $L$ over an atlas of $M$ is \emph{equivalent} to the construction of a rank $r$ vector bundle $E\to{M}$. In the very special case where the structure functions $C^a_{bc}$ are also constant, one has $E=M\times\g$ and one can separate the manifold $M$ from the abstract vector space $\g$. This is a very special situation: even as a vector bundle, $E$ may be far from trivializable.

How to obtain the algebraic structure on $E$ that, in the above particular case, would reduce to the Lie algebra structure on $\g$? Since in general we cannot separate $\g$ from $M$ inside $E$, let us thus look at $M\times\g$ also in this case, where in principle we know the action $\rho$ of the Lie algebra $\g$ on $M$, $\rho\colon\g\to\Gamma(TM)$, $\xi_a\mapsto\rho_a$. In physics we are used to associate a nilpotent odd BRST-transformation to this:
\begin{equation} \label{Q}
Q = \xi^a \rho_a^i \frac{\partial}{\partial x^i} - \frac{1}{2} C^a_{bc} \xi^b \xi^c \frac{\partial}{\partial \xi^a} \, .
\end{equation}
Here the variables $\xi^a$ are considered to be odd and, in the BRST- or BV-language are called ``ghosts''. $Q^2=0$ then follows from the above data, but also permits reciprocally to \emph{encode} the algebraic structure on $\g$ as well as its action on $M$. It is near at hand to consider the \emph{same} odd vector field \eqref{Q} also in the more generic case where the $C^a_{bc}$s are not constant. We now require that again $Q$ squares to zero. (Let us remark in parenthesis that $Q^2(x^i)=0$  as well as the contraction of the remaining condition $Q^2(\xi^a) = 0$ with $\rho_a^i$ follow already from \eqref{eq:involutive}. The additional input at this stage is mild therefore.) It is a mathematical fact \cite{Vaintrob} (cf, e.g., \cite{Gruetzmann-Strobl} for more details)
that a nilpotent vector field \eqref{Q} equips the above vector bundle $E$ with what is called a Lie algebroid structure!

As the name suggests, Lie algebroids are an infinitesimal version to Lie groupoids. We refer to the mathematics literature \cite{daSilva-Weinstein,Mackenzie,Moerdijk} for definitions and known facts about Lie algebroids and Lie groupoids.

\section{Extended Killing Equation}
Returning to Equation \eqref{hatomega}, we now recognize that the geometrical significance of $\omega^a_b$ is the one of a \emph{connection} on the bundle $E$. This permits an alternative interpretation of the main equation \eqref{gDelta} underlying the gauging. It is well-known that Killings equation ${\cal L}_vg=0$ can be rewritten in the form
\begin{equation}
v_{i;j} + v_{j;i} = 0 \, , \label{Killing}
\end{equation}
where the indices of the vector field are lowered by means of the metric $g$ and the semicolon indicates a covariant derivative with respect to the Levi-Civita connection of $g$. Any $v$ generating an isometry has to satisfy this equation.

We may reformulate the condition we found from gauging in a similar way: It is the condition on a collection of $r$ vector fields $v_a\equiv\rho_a$ to satisfy the equations \eqref{eq:involutive} and \eqref{gDelta}. Locally we can always consider this set of vector fields $v_a$ on $M$ as a single section of $E^*\times{TM}$, where $E$ is a trivial rank $r$ bundle. In fact, in the previous section we assumed that there exists a consistent gluing over local charts so as to define a not necessarily trivial rank $r$ bundle $E\to{M}$. The collection of vector fields $v_a$ now corresponds to a \emph{single} (at least locally defined) section $v\in\Gamma(E^*\otimes{TM})$ where $v=v_a^ie^a\otimes\partial_i$ with $e^a$ being a local basis in $E^*$. $g$ permits $v$ to be identified with a section of $E^*\otimes{T}^*M$. Using $\omega^a_b$ as a connection in $E$ and the Levi-Civita connection on $T^*M$, we arrive at the following compact generalization of \eqref{Killing}:
\begin{equation} \label{extendedKilling}
(\nabla v)_{\mathrm{symm}} = 0  \quad \quad \Leftrightarrow \quad \quad v_{ai;j} + v_{aj;i} = 0 \, .
\end{equation}
Certainly, an ordinary Lie algebra action of symmetries is a very particular case inside this: Then $E$ is the globally flat bundle $E=M\times\g$, we can choose $\omega^a_b=0$, and the above extended Killing equation \eqref{extendedKilling} reduces to the standard Killing equation \eqref{Killing} to be satisfied for each of the individual vector fields (or 1-forms) $v_a$ separately.

\section{Generalized gauge fields}
Since in general we cannot separate the algebraic Lie structure from the manifold $M$, one also needs to regard the scalar and gauge fields in a more unified manner. How to do this was proposed already in \cite{Bojowald-Kotov-Strobl,LAYM}: For a general Lie algebroid $E$, the fields $X^i$ and $A^a$ are in one-to-one correspondence with vector bundle morphisms
$a\colon{T}\Sigma\to{E}$ and the functional \eqref{eq:sigmagauged} gauging the $E$-orbits on $M$ is a functional of such maps $S[X^i,A^a] \equiv S[a]$. Moreover, the gauge transformations found in the second part of our Theorem are precisely one of the two options of gauge symmetries developed by independent, more mathematical considerations in \cite{Bojowald-Kotov-Strobl,LAYM,Mayer-Strobl}. In \cite{LAYM} possible purely kinetic terms for Lie algebroid Yang-Mills theories were proposed and  in \cite{Mayer-Strobl} their subsequent coupling to matter fields. The present analysis provides a complementary perspective, having started from an  ordinary and standard sigma model.

\section{Conclusion}
We have seen that by an unbiased attempt to gauge a sigma model,  not following blindly the established path of Lie groups acting as isometries on the target $M$, we are almost automatically led to the notion of Lie algebroids $E$ (or their integtating Lie groupoids). And even in the case $E=M\times\g$, we do not need the Lie algebra $\g$ to be an isometry for gauging: It is sufficient that the metric satisfies \eqref{decomp} or, if ${\cal{F}}$ is singular, the more general condition \eqref{extendedKilling}. It will be interesting to investigate this further from both sides, mathematics and physics, many new routes and problems offering themselves at this point.

Lie algebras and groups are a highly developed and important subject of mathematics and its use within physics undoubtedly indispensable. We find that in the context of gauge theories we should consider---at least also---Lie algebroids and groupoids, all the more so due to the considerable recent mathematical progress in this domain. Nature is known to have made ample use of Lie groups. It seems unlikely that Nature has restricted itself to this relatively rigid notion, not also making use of the much more flexible Lie groupoids---and this also at the level of fundamental physics. To be unravelled.

\begin{acknowledgments}
We are grateful to A.~Weinstein for valuable remarks.
\end{acknowledgments}

 \bibliography{bibtexPRL}

\begin{thebibliography}{10}%
\makeatletter
\providecommand \@ifxundefined [1]{%
 \ifx #1\undefined \expandafter \@firstoftwo
 \else \expandafter \@secondoftwo
\fi
}%
\providecommand \@ifnum [1]{%
 \ifnum #1\expandafter \@firstoftwo
 \else \expandafter \@secondoftwo
\fi
}%
\providecommand \enquote [1]{``#1''}%
\providecommand \bibnamefont  [1]{#1}%
\providecommand \bibfnamefont [1]{#1}%
\providecommand \citenamefont [1]{#1}%
\providecommand\href[0]{\@sanitize\@href}%
\providecommand\@href[1]{\endgroup\@@startlink{#1}\endgroup\@@href}%
\providecommand\@@href[1]{#1\@@endlink}%
\providecommand \@sanitize [0]{\begingroup\catcode`\&12\catcode`\#12\relax}%
\@ifxundefined \pdfoutput {\@firstoftwo}{%
 \@ifnum{\z@=\pdfoutput}{\@firstoftwo}{\@secondoftwo}%
}{%
 \providecommand\@@startlink[1]{\leavevmode\special{html:<a href="#1">}}%
 \providecommand\@@endlink[0]{\special{html:</a>}}%
}{%
 \providecommand\@@startlink[1]{%
  \leavevmode
  \pdfstartlink
   attr{/Border[0 0 1 ]/H/I/C[0 1 1]}%
   user{/Subtype/Link/A<</Type/Action/S/URI/URI(#1)>>}%
  \relax
 }%
 \providecommand\@@endlink[0]{\pdfendlink}%
}%
\providecommand \url  [0]{\begingroup\@sanitize \@url }%
\providecommand \@url [1]{\endgroup\@href {#1}{\urlprefix}}%
\providecommand \urlprefix [0]{URL }%
\providecommand \Eprint[0]{\href }%
\@ifxundefined \urlstyle {%
  \providecommand \doi [1]{doi:\discretionary{}{}{}#1}%
}{%
  \providecommand \doi [0]{doi:\discretionary{}{}{}\begingroup
  \urlstyle{rm}\Url }%
}%
\providecommand \doibase [0]{http://dx.doi.org/}%
\providecommand \Doi[1]{\href{\doibase#1}}%
\providecommand \bibAnnote [3]{%
  \BibitemShut{#1}%
  \begin{quotation}\noindent
    \textsc{Key:}\ #2\\\textsc{Annotation:}\ #3%
  \end{quotation}%
}%
\providecommand \bibAnnoteFile [2]{%
  \IfFileExists{#2}{\bibAnnote {#1} {#2} {\input{#2}}}{}%
}%
\providecommand \typeout [0]{\immediate \write \m@ne }%
\providecommand \selectlanguage [0]{\@gobble}%
\providecommand \bibinfo [0]{\@secondoftwo}%
\providecommand \bibfield [0]{\@secondoftwo}%
\providecommand \translation [1]{[#1]}%
\providecommand \BibitemOpen[0]{}%
\providecommand \bibitemStop [0]{}%
\providecommand \bibitemNoStop [0]{.\EOS\space}%
\providecommand \EOS [0]{\spacefactor3000\relax}%
\providecommand \BibitemShut [1]{\csname bibitem#1\endcsname}%
\bibitem{Note1}%
  \BibitemOpen
  \bibinfo {note} {We remark that $C^a_{bc}$ is not yet uniquely defined by
  means of \protect \textup {\hbox {\mathsurround \z@ \protect \normalfont
  (\ignorespaces \ref {eq:involutive}\unskip \@@italiccorr )}} since the set of
  $\rho _a$s does not form a basis in general. At the moment we only assume a
  smooth, consistent choice being made.}%
  \bibAnnoteFile{Stop}{Note1}%
\bibitem{Neill}%
  \BibitemOpen
  \bibfield{author}{%
  \bibinfo {author} {\bibfnamefont{B.}~\bibnamefont{O'Neill}},\ }%
  \bibfield{journal}{%
  \bibinfo {journal} {The Michigan Mathematical Journal}\ }%
  \textbf{\bibinfo {volume} {13, no. 4}},\ \bibinfo {pages} {459} (\bibinfo
  {year} {1966})%
  \bibAnnoteFile{NoStop}{Neill}%
\bibitem{Vaintrob}%
  \BibitemOpen
  \bibfield{author}{%
  \bibinfo {author} {\bibfnamefont{A.}~\bibnamefont{Vaintrob}},\ }%
  \bibfield{journal}{%
  \bibinfo {journal} {Uspekhi Matem. Nauk.}\ }%
  \textbf{\bibinfo {volume} {52(2)}},\ \bibinfo {pages} {428} (\bibinfo {year}
  {1997})%
  \bibAnnoteFile{NoStop}{Vaintrob}%
\bibitem{Gruetzmann-Strobl}%
  \BibitemOpen
  \bibfield{author}{%
  \bibinfo {author} {\bibfnamefont{M.}~\bibnamefont{Gruetzmann}}\ and\ \bibinfo
  {author} {\bibfnamefont{T.}~\bibnamefont{Strobl}},\ }%
  \enquote{\bibinfo {title} {General {Y}ang-{M}ills type gauge theories for
  p-form gauge fields: {F}rom physics to mathematics},}\  (\bibinfo {year}
  {2014})%
  \bibAnnoteFile{NoStop}{Gruetzmann-Strobl}%
\bibitem{daSilva-Weinstein}%
  \BibitemOpen
  \bibfield{author}{%
  \bibinfo {author} {\bibfnamefont{A.~C.}\ \bibnamefont{da~Silva}}\ and\
  \bibinfo {author} {\bibfnamefont{A.}~\bibnamefont{Weinstein}},\ }%
  \emph{\bibinfo {title} {Geometric Models for Noncommutative Algebras}}\
  (\bibinfo {publisher} {AMS},\ \bibinfo {address} {Providence, RI},\ \bibinfo
  {year} {1999})%
  \bibAnnoteFile{NoStop}{daSilva-Weinstein}%
\bibitem{Mackenzie}%
  \BibitemOpen
  \bibfield{author}{%
  \bibinfo {author} {\bibfnamefont{K.~C.~H.}\ \bibnamefont{Mackenzie}},\ }%
  \emph{\bibinfo {title} {Theory of Lie Groupoids and Lie Algebroids}}\
  (\bibinfo {publisher} {Cambridge University Press},\ \bibinfo {address}
  {Cambridge},\ \bibinfo {year} {2005})%
  \bibAnnoteFile{NoStop}{Mackenzie}%
\bibitem{Moerdijk}%
  \BibitemOpen
  \bibfield{author}{%
  \bibinfo {author} {\bibfnamefont{I.}~\bibnamefont{Moerdijk}}\ and\ \bibinfo
  {author} {\bibfnamefont{J.}~\bibnamefont{Mrcun}},\ }%
  \emph{\bibinfo {title} {Introduction to foliations and Lie groupoids}}\
  (\bibinfo {publisher} {Cambridge University Press},\ \bibinfo {address}
  {Cambridge},\ \bibinfo {year} {2003})%
  \bibAnnoteFile{NoStop}{Moerdijk}%
\bibitem{Bojowald-Kotov-Strobl}%
  \BibitemOpen
  \bibfield{author}{%
  \bibinfo {author} {\bibfnamefont{M.}~\bibnamefont{Bojowald}}, \bibinfo
  {author} {\bibfnamefont{A.}~\bibnamefont{Kotov}},\ and\ \bibinfo {author}
  {\bibfnamefont{T.}~\bibnamefont{Strobl}},\ }%
  \bibfield{journal}{%
  \bibinfo {journal} {J.Geom.Phys.}\ }%
  \textbf{\bibinfo {volume} {54}},\ \bibinfo {pages} {400} (\bibinfo {year}
  {2004})%
  \bibAnnoteFile{NoStop}{Bojowald-Kotov-Strobl}%
\bibitem{LAYM}%
  \BibitemOpen
  \bibfield{author}{%
  \bibinfo {author} {\bibfnamefont{T.}~\bibnamefont{Strobl}},\ }%
  \bibfield{journal}{%
  \bibinfo {journal} {Phys. Rev. Lett.}\ }%
  \textbf{\bibinfo {volume} {93}},\ \bibinfo {pages} {211601} (\bibinfo {year}
  {2004})%
  \bibAnnoteFile{NoStop}{LAYM}%
\bibitem{Mayer-Strobl}%
  \BibitemOpen
  \bibfield{author}{%
  \bibinfo {author} {\bibfnamefont{C.}~\bibnamefont{Mayer}}\ and\ \bibinfo
  {author} {\bibfnamefont{T.}~\bibnamefont{Strobl}},\ }%
  \bibfield{journal}{%
  \bibinfo {journal} {J.Geom.Phys.}\ }%
  \textbf{\bibinfo {volume} {59}},\ \bibinfo {pages} {1613} (\bibinfo {year}
  {2009})%
  \bibAnnoteFile{NoStop}{Mayer-Strobl}%
\end{thebibliography}%
\end{document}